\documentclass[useAMS,usenatbib,referee]{mn2e}
\usepackage{graphicx, amssymb}

\begin{document}

%
%

\title[EBAS --- a new Eclipsing Binary Automated Solver with EBOP]
{Automated analysis of eclipsing binary lightcurves. I. EBAS --- a new
Eclipsing Binary Automated Solver with EBOP}

\author[O. Tamuz T. Mazeh \& P. North]{O. Tamuz$^1$\thanks{E-mail:
omert@wise.tau.ac.il} T. Mazeh$^1$ and P. North$^2$\\$^1$School of
Physics and Astronomy, Raymond and Beverly Sackler Faculty of Exact
Sciences, Tel Aviv University, Tel Aviv, Israel\\$^2$Ecole
Polytechnique F\'ed\'erale de Lausanne (EPFL), Laboratoire d'Astrophysique,
Observatoire,\\ CH-1290 Sauverny, Switzerland}

\maketitle

\begin{abstract}

We present a new algorithm --- Eclipsing Binary Automated Solver (EBAS),
to analyse lightcurves of eclipsing binaries. The algorithm is
designed to analyse large numbers of lightcurves, and is
therefore based on the relatively fast EBOP code.  To facilitate the
search for the best solution, EBAS uses two parameter
transformations. Instead of the radii of the two stellar components,
EBAS uses the sum of radii and their ratio, while the inclination
is transformed into the impact parameter.  To replace human visual
assessment, we introduce a new 'alarm' goodness-of-fit statistic that
takes into account correlation between neighbouring residuals.  We
perform extensive tests and simulations that show that our algorithm
converges well, finds a good set of parameters and provides reasonable
error estimation.
\end{abstract}


\section{Introduction}

The advent of large CCDs for the use of astronomical studies has
driven a number of photometric surveys that have produced
unprecedentedly large sets of lightcurves of eclipsing binaries
\citep[e.g.,][]{alcocketal97}. The commonly used interactive way of
finding the set of parameters that best fit an eclipsing binary
lightcurve utilizes human guess for the starting point of the
iteration, and further human decisions along the converging iteration
\citep[e.g.,][]{ribas2000}. Such a process is not always repeatable,
and is impractical when it comes to the large set of lightcurves at
hand.

The OGLE project, for example, yielded a huge photometric dataset of
the SMC \citep{udalski1998} and the LMC \citep{udalski2000}, which
includes a few thousand eclipsing binary lightcurves
\citep{lukas2003}. This dataset allows for the first time a
statistical analysis of the population of short-period binaries in
another galaxy. A first effort in this direction was performed by
\citet[][hereafter NZ03]{NZ03}, who derived the orbital elements and
stellar parameters of 153 eclipsing binaries in the SMC in order to
study the statistical dependence of the eccentricity of the binaries
on their separation. In a following study, \citet[][hereafter
NZ04]{NZ04} analyzed another sample of 509 lightcurves selected from
the 2580 eclipsing binaries discovered in the LMC by the OGLE team
\citep[][]{lukas2003}. However, the OGLE LMC data contain many more
eclipsing binary lightcurves. An automated algorithm would have made
an analysis of the whole sample possible.

To meet the need for an algorithm that can handle a large number of
lightcurves we developed EBAS ---
Eclipsing Binary Automated Solver, which is a completely automatic
scheme that derives the orbital parameters of eclipsing binaries. Such
an algorithm can be of use for the OGLE lightcurves, as we do in
the next paper, and for the data of the many other large photometric
surveys that came out in the last few years (e.g., EROS, MACHO,
DIRECT, MOA). EBAS is specifically designed to quickly solve large
numbers of lightcurves with $S/N$ typical of such surveys.

\citet[][hereafter WW1]{WW1} have already developed an automatic
scheme to analyze the OGLE lightcurves detected in the SMC, in
order to find eclipsing binaries suitable for distance
measurements. However, whereas WW1 used the Wilson-Devinney (=WD)
code, EBAS uses the EBOP code, which is admittedly less accurate than
the WD code, but much simpler and faster. We used the EBOP
\citep[][]{PE81,etzel81} subroutines that generate an eclipsing binary
lightcurve for a given set of orbital elements and stellar parameters,
and rewrote a {\it fully automated} iterative code that finds the best
parameters to fit the observed lightcurve.

 As EBAS uses extensively
the lightcurve generator for each system, we preferred EBOP over the
WD code.

At the last stages of writing this paper another study with an
automated lightcurve fitter --- Detached Eclipsing Binary Lightcurve
(DEBiL), was published \citep{devor2005}.  DEBiL was constructed to be
quick and simple, and therefore has its own lightcurve generator,
which does not account for stellar deformation and reflection
effects. This makes it particularly suitable for detached binaries. The
complexity of the EBOP lightcurve generator is in between DEBiL and the
automated WD code of WW1.

To facilitate the search for the global minimum in the convolved parameter
space, EBAS performs two parameter transformations. Instead of the
radii of the two stellar components of the binary system, measured in
terms of the binary separation, EBAS uses two other parameters, the
sum of radii (the sum of the two relative radii), and their
ratio. Instead of the inclination we use the impact parameter --- the
projected distance between the centres of the two stars during the
primary eclipse, measured in terms of the sum of radii.

During the development of EBAS we found that some solutions with low
$\chi^2$ could easily be classified as flawed by visual inspection
that revealed correlation between neighbouring residuals.  We have
therefore developed a new 'alarm' statistic, $\mathcal{A}$, to replace
human inspection of the residuals. EBAS uses this statistic to decide
automatically whether a solution is satisfactory.

The EBAS strategy consists of three stages. First, EBAS finds a good
initial guess by a combination of grid searches, gradient descents and
geometrical analysis of the lightcurve. Next, EBAS searches for the
global minimum by a simulated annealing algorithm. Finally, we asses
the quality of the solution with the new 'alarm' statistic, and if
necessary, perform further minimum searches.

To check our new algorithm we ran many simulations which demonstrated
that the automated code does find the correct values of the orbital
parameters. We also used simulations to estimate the error induced by
two of our simplifying assumptions, namely mass ratio of unity and
negligible third light. We then checked the code against the results
of NZ04, and found that our code performed as well as their
interactive scheme, except for very few systems. Finally, we checked
our code against four LMC eclipsing binaries that were solved by
\citet{gonzalez05} using photometry and radial-velocity data.

Section~\ref{sec:parameters} presents the EBAS parameters and compares
them with the EBOP ones.  Section~\ref{sec:search} details how the
algorithm finds the global minimum of the $\chi^2$ function, and
Section~\ref{sec:alarm} describes our new alarm statistic. In
Section~\ref{sec:simulation} and Section~\ref{sec:discussion} we check
and discuss the performance of EBAS.

\section{The EBAS Parameters}
\label{sec:parameters}

EBAS is based on the the EBOP code \citep[][]{PE81,etzel81}, which
consists of two main components. The first component generates a
lightcurve for a given set of orbital elements and stellar parameters,
while the second finds the parameters that best fit the observational
data. We only used the lightcurve generator, and wrote our own code to
search for the best-fit elements.

Like all other model fitting algorithms, EBAS searches for the global
minimum of the $\chi^2$ function in the space spanned by the
parameters of the model. The natural parameters of an eclipsing binary
model include the radii of the two stars relative to the orbital
semi-major axis, the relative surface brightness of the two stars,
$J_s$, the orbital parameters of the system, $P$, $T_0$, $e$ and
$\omega$, and some parameters that characterize the shape of the two
stars and the light distribution over their surface, such as limb and
gravity darkening coefficients.

Finding the global minimum can be quite difficult, because the
parameter space of the model is complex and convoluted,
causing the $\chi^2$ function to have many local minima. Therefore,
the choice of parameters might be of particular importance, as a
change of variables can substantially modify the topography of the
goodness-of-fit function. Smart choices of the variables can allow for
a better initial guess of the parameter values, as well as more
efficient performance of the minimization algorithm.

This approach was already recognized by the writers of EBOP
\citep[][]{etzel81} who transformed the variables $e$ (eccentricity)
and $\omega$ (longitude of periastron), which have a clear Keplerian
meaning, into $e\cos\omega$ and $e\sin\omega$. This approach is
beneficial because $e\cos\omega$ corresponds closely to the difference
in phase between the primary and secondary eclipses, the two most
prominent features of the lightcurve.

Following this approach, we chose to transform the two most
fundamental parameters of the stellar components of the binary system
--- the two relative radii, $r_p=R_p/a$ and $r_s=R_s/a$, where $R_p$
and $R_s$ are the radii of the primary and the secondary and $a$ is
the orbital semi-major axis. Instead, we used the sum of radii
$r_t=(R_p+R_s)/a$ and $k=R_s/R_p$, because the sum of radii can be
well determined from the lightcurve, much better than $r_p$ or
$r_s$. With the same reasoning we chose to parameterize the lightcurve
by the impact parameter, $x$, which measures the projected distance
between the centres of the two stars in the middle of the primary
eclipse (i.e. at phase zero), in terms of the sum of radii $r_t$:
\begin{equation}
x=\frac{\cos i}{r_t}\frac{1-e^2}{1+e\sin\omega}.
\end{equation}
Thus, $x=0$ when $i=\pi/2$ and $x=1$ when the components are
grazing but not yet eclipsing. We found that the impact parameter is
directly associated with the shape of the two eclipses, and can therefore
be determined much better than $i$, the more conventional
parameter.

The EBOP lightcurve generator models the stellar shapes by simple
biaxial and similar ellipsoids, instead of calculating the actual
shapes of the two binary components. This means that systems with
components which suffer from strong tidal deformation are poorly
modelled. Furthermore, unphysical parameters sets, with stars larger
than their Roche lobes, for example, are permissible by EBOP. We
therefore limit ourselves to stars which are likely to be
significantly smaller than their Roche lobes. 

Using the formula in
\citet{Egg1983}:
\begin{equation}
R_{RL}/a=\frac{0.49\,q^{2/3}}{0.6\, q^{2/3}+\ln(1+q^{1/3})} \ ,
\end{equation}
which reduces for $q=1$ (see below) to $R_{RL}/a=0.379$, we do not
accept solutions with $(R_p+R_s)/a>0.65(1-e\cos\omega)$.

The bolometric reflection of the two stars are varied in EBAS by $A_p$
and $A_s$. When $A_p=1$, the primary star reflects all the light cast
on it by the secondary. Together with the tidal distortion of the two
components, which is mainly determined by the mass ratio of the two
stars, the reflection coefficients $A_p$ and $A_s$ determine the light
variability of the system outside the eclipses. Note, however, that
the EBOP manual \citep[][]{etzel81} stresses that the model at
the basis of the programme is a crude approximation to the real
variability outside the eclipses. Therefore, the EBOP manual warns
against the reliability of the reflection parameters derived by the
code. Nevertheless, we decided to vary $A_p$ and $A_s$, in order to
fit the out-of-eclipse variability, even with an improbable (but not
physically impossible) model for some cases.  By doing that we could
allow the algorithm to find the value of the other parameters that
best fit the actual shape of the two eclipses. The reflection
coefficients should be viewed as two extra free parameters of the fit,
and not as physical quantities determined by the lightcurve.

The EBOP manual defines the primary as the component eclipsed at phase
$0$, probably because the general practice assigns this phase to
the deeper eclipse. However, this definition leaves the freedom
to change the zero phase of the lightcurve and therefore interchange
between the primary and the secondary in the resulting solution. To
prevent such ambiguity, we chose the primary as being the star with
the higher surface brightness. Consequently, if the solution showed
$J_s>1$, we switched the components. Accordingly, in EBAS the primary
is the star with the higher surface brightness, and not necessarily
the larger star. In special cases with eccentric orbits, the primary
might not even be the star which is eclipsed at the deeper eclipse.

All relevant parameters of EBAS in this work are listed in
Table~\ref{tab:parameters_EBAS}.

\begin{table*}
\label{tab:parameters_EBAS}
 \centering
 \begin{minipage}{140mm}
  \caption{EBAS parameters}
  \begin{tabular}{@{}lll@{}}
   Symbol & Parameter\\
 \hline
 $J_s$          & Surface brightness ratio (secondary/primary)  \\
 $r_t$          & Fractional sum of radii \\
 $k$            & Ratio of radii (secondary/primary) \\
 $x$            & Fractional impact parameter \\
 $e\cos\omega$  & Eccentricity times the cosine of the longitude of periastron \\
 $e\sin\omega$  & Eccentricity times the sine of the longitude of periastron \\
 $A_p$          & Primary bolometric reflection coefficient \\
 $A_s$          & Secondary bolometric reflection coefficient \\
 $T_0$          & Time of primary eclipse  \\
 $P$            & Period\\
\hline
\end{tabular}
\end{minipage}
\end{table*}

The present embodiment of EBAS is aimed at solving lightcurves from
surveys such as the OGLE LMC and SMC studies. Many of these lightcurves
have low $S/N$ ratio, and therefore the mass ratio and limb and gravity
darkening can not be found reliably. We therefore decided not to vary
these parameters, and adopted here a unity value for the value of the
mass ratio, and 0.18 and 0.35 for the values of limb and gravity
darkening coefficients, respectively, for both the primary and the
secondary.  The last two values are suitable for early-type
stars, which form the major part of the OGLE eclipsing binary
sample of the LMC and SMC. This does not mean that EBAS (through EBOP)
does not model tidal distortion and limb and gravity darkening,
but only that in all cases shown in this paper, optimization is not
performed on these parameters. In other implementations of EBAS, more
parameters could be varied.

Table~\ref{tab:EBOP} brings the full list of EBOP parameters we use to
generate the lightcurves, as they appear in the EBOP manual
\citep[][]{etzel81}, which describes them in detail.  The table
also explains how to derive the EBOP parameters which are not used as
EBAS parameters in the present version. The $L_p$ and $L_s$ terms in
the formulae for $A_p$ and $A_s$ are the EBOP parameters for the
contribution of the primary and secondary to the total light of the
system. See the EBOP manual \citep[][]{etzel81} for more detail.

\begin{table*}
\label{tab:EBOP}
 \centering
 \begin{minipage}{140mm}
  \caption{EBOP lightcurve generator parameters}
  \begin{tabular}{@{}rll@{}}
   Symbol & Parameter & Calculation from EBAS parameters\\
 \hline
 $J_s$          & Surface brightness ratio (secondary/primary) &\\
 $r_p$          & Fractional radius of primary & $r_tk$\\
 $k$            & Ratio of radii (secondary/primary) &\\
 $u_p$       & Limb darkening coefficient of primary & constant: 0.18\\
 $u_s$       & Limb darkening coefficient of secondary & constant: 0.18\\
 $i$            & Inclination & $\cos i=r_tx\frac{1+e\sin\omega}{1-e^2}$ \\
 $e\cos\omega$  & Eccentricity and longitude of periastron &\\
 $e\sin\omega$  & Eccentricity and longitude of periastron &\\
 $y_p$     & Gravity darkening coefficient of primary & constant: 0.35\\
 $y_s$     & Gravity darkening coefficient of secondary & constant: 0.35\\
 $S_p$     & Reflected light from primary & $0.4L_s r_p^2 A_p$\\
 $S_s$     & Reflected light from secondary & $0.4L_p r_s^2 A_s$\\
 $q$            & Mass ratio & constant: 1\\
 $t$            & Tidal lead/lag angle & constant: 0\\
 $L_3$          & Third light (blending) & constant: 0 \\
 $T_0$          & Time of primary eclipse  &\\
 $SFACT$        & Luminosity scaling factor & Linear factor - solved analytically\\
\hline
\end{tabular}
\end{minipage}
\end{table*}

Note that two more EBOP parameters are not varied in the present
version of EBAS: the tidal lead/lag angle, $t$, and the light fraction
of a possible third star, $L_3$.  Both elements are put to zero.  We
estimate the implication of the latter assumption in
Section~\ref{sec:simulation}. On the other hand, the orbital period,
which is a fitted parameter of EBAS, is not included in the EBOP list
of parameters. EBOP assumes the period is known and therefore all
observing timings are given in terms of the orbital phases. EBAS
partly follows this approach and does not perform an initial search
for the best period. However, EBAS does try to improve on the initial
guess of the period after solving for all the other parameters.  For
this purpose, the timings of the observational data points need to be
given, and not only their phases. Note that this approach requires
the original guess for the period to be close to the real one.

\section{Searching for the $\chi^2$ minimum}
\label{sec:search}

The search for the global $\chi^2$ minimum is performed in two
stages. We first find a good initial guess, and then use a simulated
annealing algorithm to find the global minimum.  While the first stage
is merely aimed at finding an initial guess for the next stage, in
most cases it already converges to a very good solution.

The initial guess search starts by fitting the lightcurve with a small
number of parameters, and then adding more and more parameters, till
the full set of parameters is reached.  The smaller number of
parameters in the first steps makes this process converge quickly and
efficiently. The values of the parameters as determined in each step
are very preliminary, and are useful {\it only} to facilitate the next
steps. This is done in five steps:

\begin{enumerate}

\item Finding $T_0$ by identifying the primary eclipse and the phase
of its centre.

\item Fitting a lightcurve to the {\it primary eclipse only}, with
$r_t$, $k$, $x$, and $T_0$ as free parameters.

\item Finding $e\cos\omega$ by determining the phase of the centre of
  the secondary eclipse.

\item Fitting the whole lightcurve with two additional parameters,
$J_s$ and $e\sin\omega$.

\item Finding the nearest $\chi^2$ local minimum, allowing all parameters
to vary.

\end{enumerate}

The searches for the best parameters are first done over a grid of the
pertinent parameters, followed by optimization with the
Levenberg-Marquardt algorithm \citep{marquardt}, implemented by the
{\it matlab} minimization routine {\it lsqnonlin}.

Having found an initial guess, EBAS proceeds to improve the model by
using a variation of the {\it matlab} downhill simplex routine {\it
fminsearch}. Following \cite{NR}, we combine this procedure with the
simulated annealing technique, allowing it to 'roll' uphill
occasionally and leave local minima.

\begin{figure*}
 \includegraphics{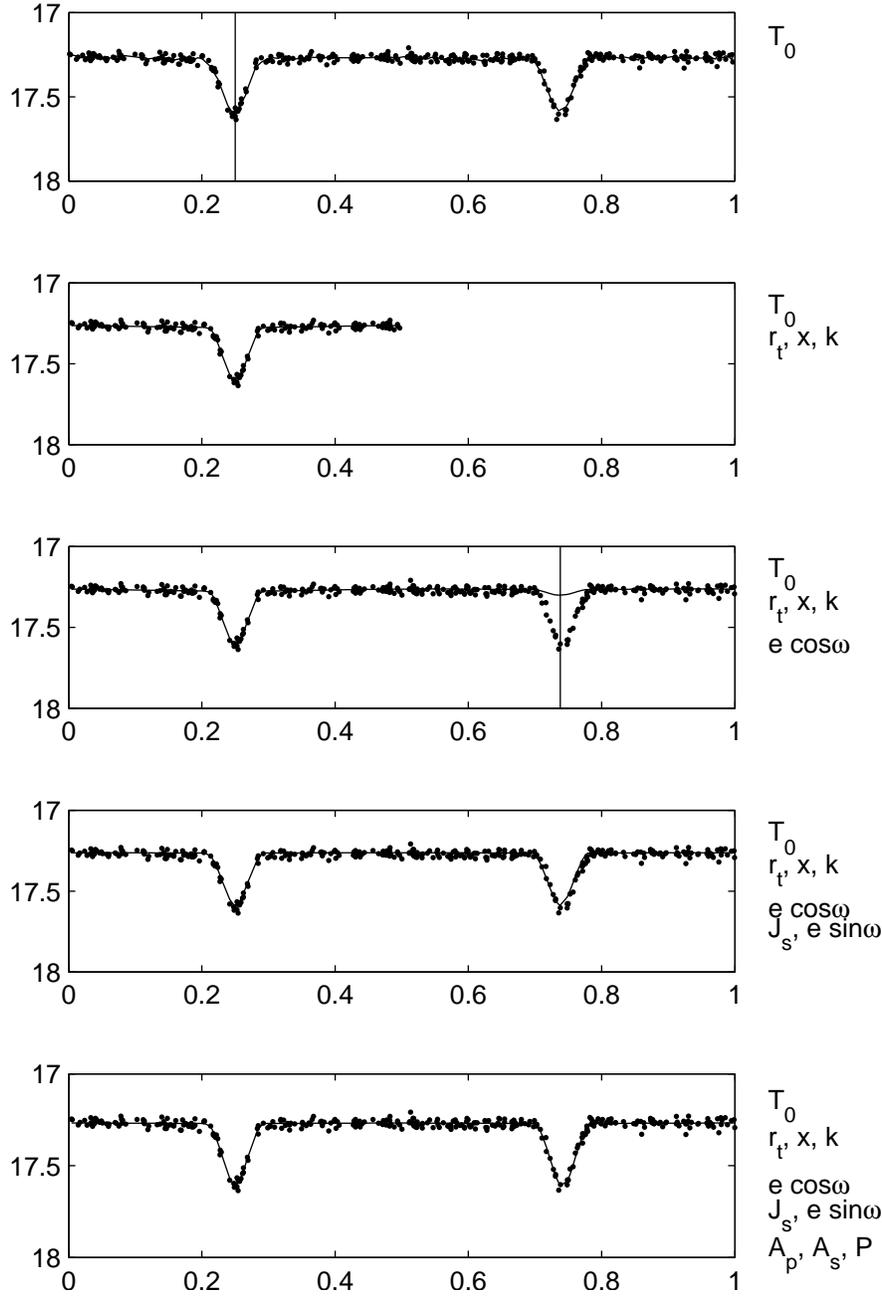}
\caption{The five steps of obtaining the best initial guess for
OGLE053312.82-700702.5. The values of the parameters in each step are
given in Table~\ref{tab:steps_initial}. The line in the first panel
is a smoothing of the data, performed by a running mean smoothing algorithm. 
The rest of the lines are EBAS
models in the different steps of the algorithm. Note that the bottom panel
presents the best initial guess and not the final solution. The
vertical lines in the first and the third panels are EBAS best estimate
for the centres of the two eclipses.}
\label{fig:steps_initial}
\end{figure*}

Fig~\ref{fig:steps_initial} demonstrates the EBAS procedure by showing the
five steps of finding the initial guess and the final solution of
OGLE053312.82-700702.5. The values of the parameters in each of the
five steps are given in Table~\ref{tab:steps_initial}. The last column
of the table brings the $\chi^2$ value of the solution. This is done
only for the steps which fit the whole lightcurve.

\begin{table*}
\label{tab:steps_initial}
\centering
\begin{minipage}{140mm}
\caption{The five steps of obtaining the initial guess for
OGLE053312.82-700702.5.}
\begin{tabular}{@{}lrrrrrrrrrrr@{}}
Stage & $T_0 \ \ \ $ & $r_t$ \ \ & $x \ \ \ $ & $k$ \ \ \ \,&
$e\cos\omega$ & $J_s$ \ \ & $e\sin\omega$ \,&
$A_p$ \ & $A_s\ $ & $P$\,\,\,& $\chi^2\ $\\
\hline
1 &   729.87\\
2 &   729.85 & 0.2805 & 0.4911 & 0.6810 \\
3 &   729.85 & 0.2805 & 0.4911 & 0.6810 & -0.0184 & & & & & & 2103.5\\
4 &   729.85 & 0.2805 & 0.4000 & 1.0000 & -0.0184 & 1.0000 & 0.0000 & & &
& 323.9\\
5 &   729.85 & 0.2771 & 0.3806 & 0.9996 & -0.0136 & 1.0213 &  -0.0003 &
0.9834 & 0.9998 & 5.394410 & 267.6\\
\hline
final &   729.85 & 0.2697 & 0.3185 & 1.5087 & -0.0134 & 1.0414 &  0.0143 &
0.3948 & 0.9927 & 5.394382 & 263.9\\
\hline
\end{tabular}
\end{minipage}
\end{table*}

To estimate the uncertainties of the derived parameters, EBAS uses the
Monte-Carlo bootstrap method, as described in \cite{NR}. For each
solution, we generated a set of 25 simulated lightcurves by using
the values of the model at the original data points, with added normally
distributed noise. The amplitude of the noise is chosen to equal the
uncertainty of the data points. EBAS then proceeds to solve each of
these lightcurves, using the simulated (''true'') values of $r_t$, $T_0$, $P$
and $e\cos\omega$ as initial guesses. EBAS sets the error of each
parameter to be the standard deviation of its values in the sample of
generated solutions.  Section~\ref{sec:simulation} analyses the performance
of EBAS and finds that its error estimation is correct to a factor of
about 2.

\section{A new 'alarm' statistic to assess the solution goodness-of-fit}
\label{sec:alarm}

\begin{figure*}
 \includegraphics{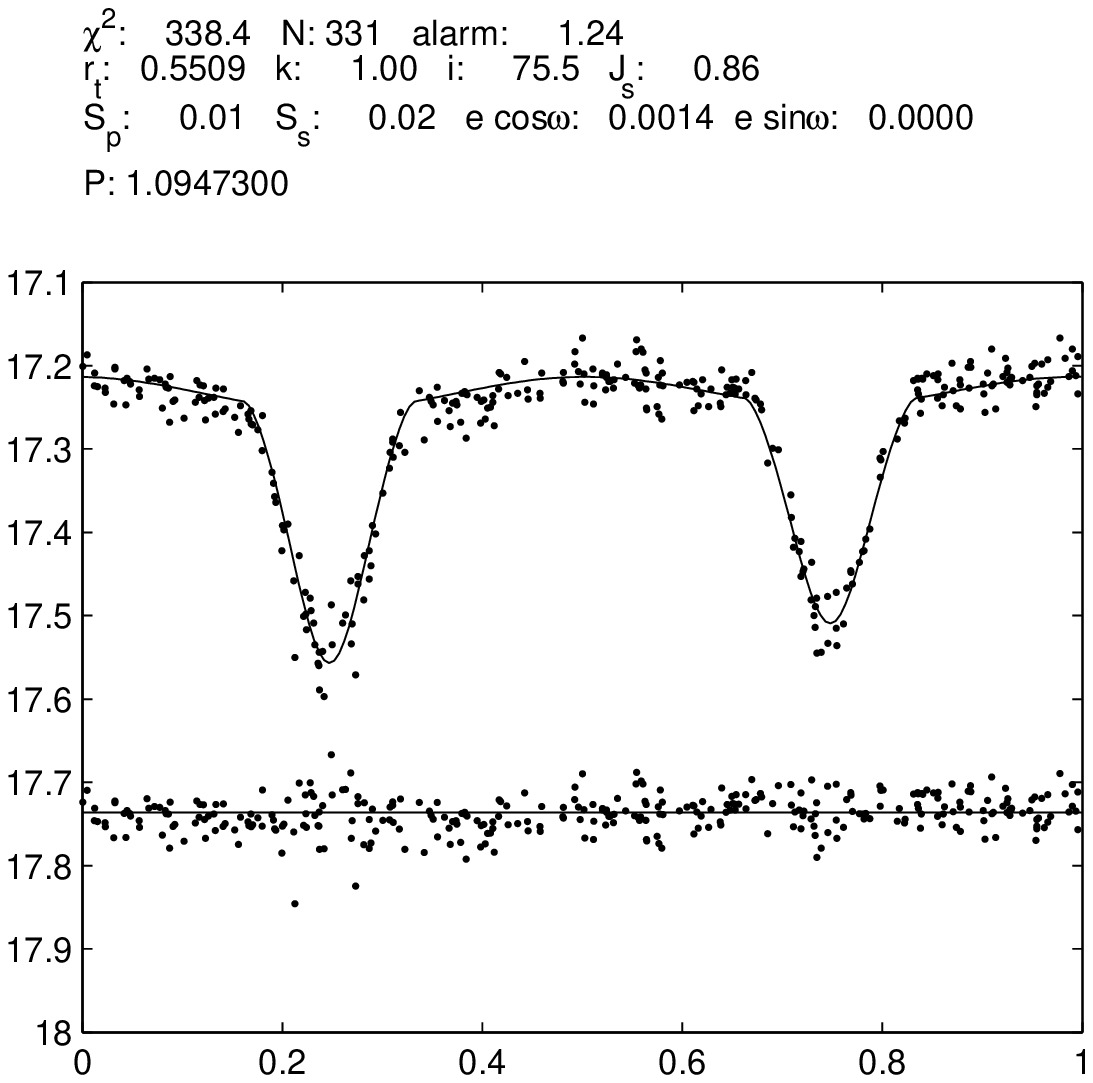}
\caption{Lightcurve, solution and elements for OGLE051331.74-691853.5,
  as derived by NZ04. The solution is not optimal, as visual scrutiny
  of the edges of the eclipses may reveal.}
 \label{fig:highalarm}
\end{figure*}

During the development of EBAS we found that some solutions with low
$\chi^2$ might be unsatisfactory.  Fig~\ref{fig:highalarm} presents
such a system solution, OGLE051331.74-691853.5, obtained manually by
NZ04. While the value of $\chi^2$ is reasonable,
the model deviates from the observations at the edges of the
eclipses, as a visual inspection of the
residuals, plotted as a function of phase, can reveal. This case shows
that the $\chi^2$ statistic, while being the unchallenged
goodness-of-fit indicator, can be low even for solutions which are not
quite satisfactory.  For such cases, human interaction is needed to
improve the fit, or to otherwise decree the solution
unsatisfactory.  In order to allow an automated approach, an automatic
algorithm must replace human evaluation.

We therefore defined a new estimator which is
sensitive to the correlation between adjacent residuals of the
measurements relative to the model. This feature is in contrast to the
behaviour of the $\chi^2$ function, which measures the sum of the
squares of the residuals, but is not sensitive to the signs of the
different residuals and their order. For an estimator to be sensitive
to the number of {\it consecutive} residuals with the same sign, one
might use some kind of run test \citep[e.g.,][]{kanji}. In such a
test, the whole lightcurve is divided into separate sequential runs,
where a 'run' is defined as a maximal series of consecutive residuals
(in the folded lightcurve) with the same sign.  For example, if the
residuals are $\{1, 2, 1, -3, -4, 5, -2, -3\}$ (written in order of
increasing phase) the four runs would be $\left\{\{1, 2, 1\}, \{-3,
-4\}, \{5\}, \{-2, -3\}\right\}$.  Long runs might indicate that the
residuals are not randomly distributed. For example, in Fig~\ref{fig:highalarm}
a run of 13 negative residuals exists around phase $0.4$, and a
run of 17 positive residuals exists around phase $0.65$.

Different approaches for residual diagnostics based on run
tests may be found in the literature \citep[e.g.,][]{kanji}. Lin's
cumulative residuals \citep[][]{Lin2000} is one example. We chose to
define a new estimator which is sensitive both to the length of the
runs and to the magnitude of the residuals, in units of their
uncertainties.

Denoting by $k_i$ the number of residuals in the $i$-th run, we define
the 'alarm' $\mathcal A$ as:

\begin{equation}
\mathcal{A}=\frac{1}{\chi^2}\sum_{i=1}^M{\left(\frac{r_{i,1}}{\sigma_{i,1}}
                                  + \frac{r_{i,2}}{\sigma_{i,2}}
                                  +\cdots+
                                  \frac{r_{i,k_i}}{\sigma_{i,k_i}}
                                  \right)^2}
                                  -(1+\frac{4}{\pi})\ \ ,
\end{equation}
where $r_{i,j}$ is the residual of the $j$-th measurement of the
$i$-th run and $\sigma_{i,j}$ is its uncertainty. The sum is over all
the measurements in a run and then over the $M$ runs.  The $\chi^2$ is
the known function:

\begin{equation}
\chi^2=\sum_{i=1}^N{\left( \frac{r_i}{\sigma_i} \right)^2} \ \ ,
\label{chi2}
\end{equation}
where the sum is over all $N$ observations.  Dividing by $\chi^2$
assures that, in contrast to $\chi^2$ itself, $\mathcal A$ is not
sensitive to a systematic overestimation or underestimation of the
uncertainties.

It is easy to see that $\mathcal A$ is minimal when the residuals
alternate between positive and negative values, and that long runs
with large residuals increase its value. The minimal value of the
summation is exactly $\chi^2$, and therefore the minimal value of
$\mathcal A$ is $-4/\pi$.

For $N$ {\it uncorrelated} Gaussian residuals, the expectation value
for $\mathcal A$ can be calculated under the assumptions that
$\chi^2=N$, and that $N$ is large enough to make the length of the
runs be distributed geometrically for all practical
purposes. According to this calculation, the expectation value of
$\mathcal A$ as defined above vanishes.

\begin{figure*}
 \includegraphics{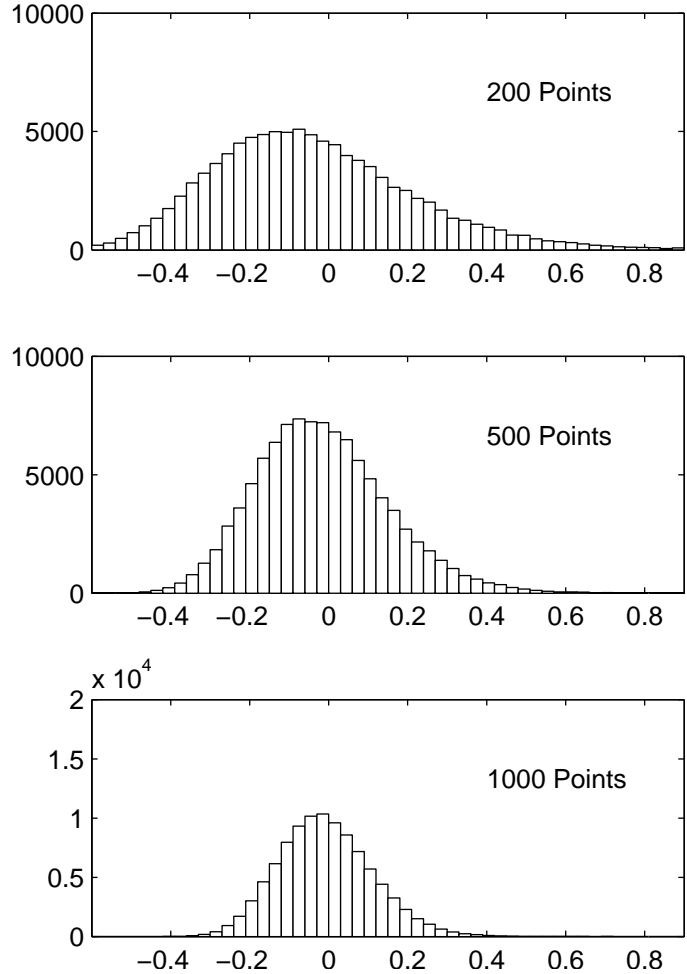} 
 \caption{The distribution of $\mathcal A$ for normally distributed
  200, 500 and 1000 random points.}
 \label{fig:alarm_hists}
\end{figure*}

To explore the behaviour of the new statistic we simulated residuals
of normal random noise composed of 200, 500 and 1000 points, each of
which for 100,000 times, and plotted in Fig~\ref{fig:alarm_hists}
histograms of the $\mathcal A$ values. The solution of
Fig~\ref{fig:highalarm} has indeed an ${\mathcal A}$ value of $1.24$,
which is too high, as can be seen in Fig~\ref{fig:alarm_hists}.

When a solution shows high $\mathcal A$, EBAS performs additional
simulated annealing searches with different initial guesses. In most
cases, a few iterations that start in the parameter space not far away
from the previously found minimum are sufficient to find a
substantially better minimum. We stop this process when EBAS finds a
new solution with low enough $\mathcal A$. If this approach does not
lead to a solution with low enough $\mathcal A$, EBAS calls for visual
inspection, and manually initialized optimization may be
attempted. Our experience with the OGLE LMC data indicated that some
systems simply cannot be modelled by the EBOP subroutines, either
because the lightcurve is not of an eclipsing binary, or because EBOP
is insufficiently accurate to model the light modulation.

\begin{figure*}
 \includegraphics{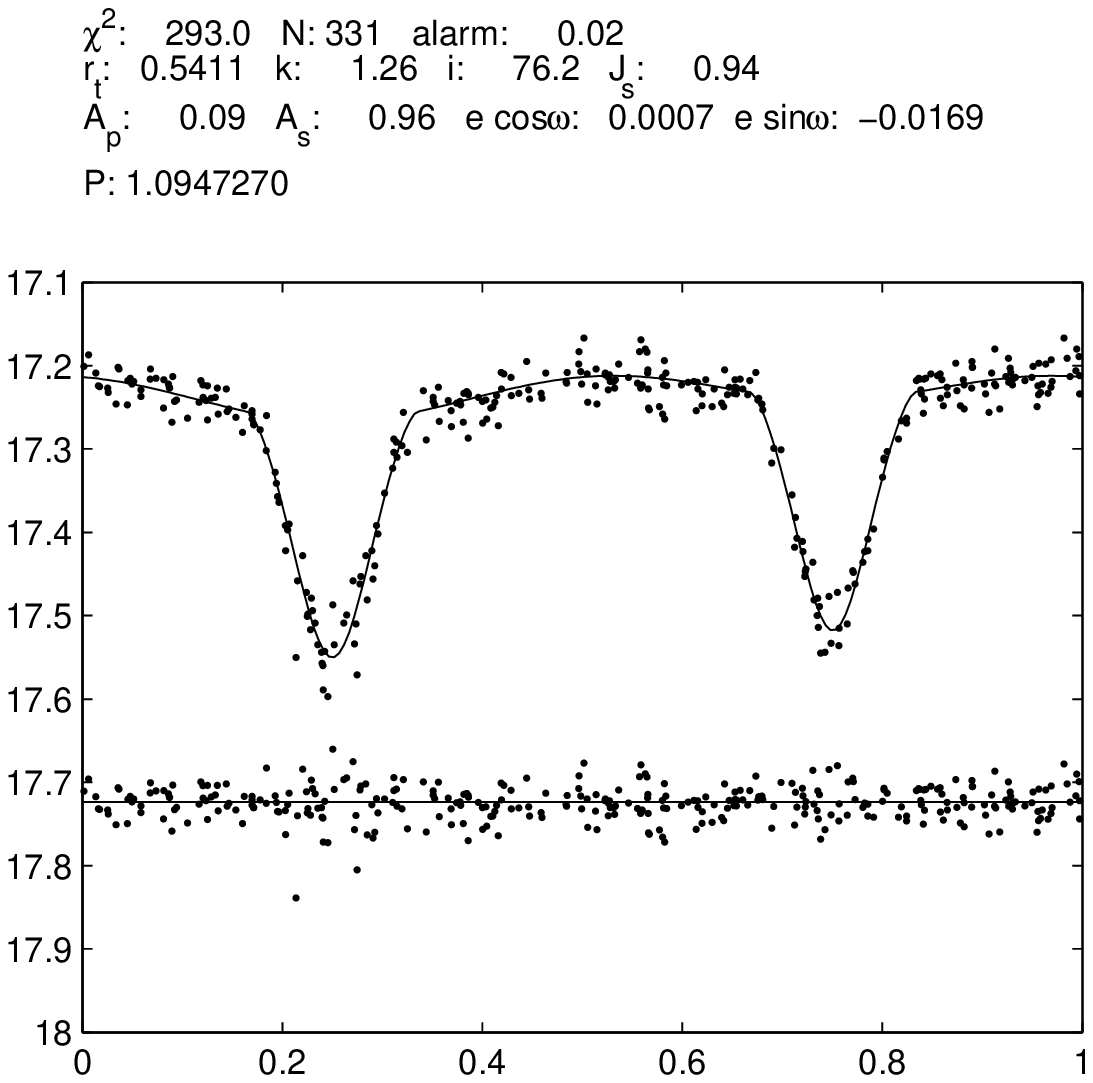}
 \caption{An improved solution relative to the one of
 Fig~\ref{fig:highalarm},  with lower $\chi^2$ and $\mathcal
 A$.}
 \label{fig:lowalarm}
\end{figure*}

Fig~\ref{fig:lowalarm} shows the lightcurve of Fig~\ref{fig:highalarm}
with its EBAS solution. Clearly, the code found a model with lower
$\chi^2$ and better $\mathcal A$ value of $0.02$.

\section{Testing the algorithm}
\label{sec:simulation}

To check the reliability of EBAS when applied to OGLE-like data, we
performed a few tests. We analyzed a large sample of {\it simulated}
lightcurves of eclipsing binaries, checked the obtained $\chi^2$
against the inserted noise, and examined the derived elements and
their uncertainties versus the correct elements. The advantage of a
simulated sample of lightcurves is the knowledge of the ``true''
elements, a feature that is missing, unfortunately, in real data.  We
also used simulations to estimate the sensitivity of the EBAS results
to the assumption that there is no contribution of light from a third
star, and to the assumption that the mass ratio is unity.  We then
compared the parameters derived by EBAS for real 509 OGLE LMC systems
with the elements obtained manually by NZ04 with the EBOP code. The
goal of this comparison was to find out how well EBAS performs as
compared with manual finding of the elements with the same
code. Finally, we compared our results with the recently derived
elements of four eclipsing binaries in the LMC by \citet[][hereafter
GOMoM05]{gonzalez05}, who analysed OGLE and MACHO photometry and a few
radial-velocity measurements. The goal of this comparison was to
compare the elements found by EBAS with elements found by using extra
information on the same systems.  This comparison is of particular
interest because GOMoM05 used for their analysis not only lightcurves
in three passbands, but also radial velocities, and they interpreted
their data with the more sophisticated WD code.

\subsection{Simulated lightcurves --- comparison with the ``true'' elements}

To check EBAS against simulated lightcurves we generated a sample of
lightcurves with the EBOP subroutines and solved them with EBAS. To
obtain an OGLE-like sample, the elements were taken from the NZ04 set
of solutions for the OGLE LMC data, with $k$, the ratio of radii,
chosen randomly from a uniform distribution between $0.5$ and $1$ (the
NZ04 solutions had $k=1$, except for the relatively few systems with
clearly total eclipses).  For each simulated system we created a
lightcurve with the original OGLE observational timings, and added
random Gaussian noise, with an amplitude equal to the rms of the
actual residuals relative to the NZ04 solution of that system. In
total, 423 simulated lightcurves were created and solved.

\begin{figure*}
\includegraphics{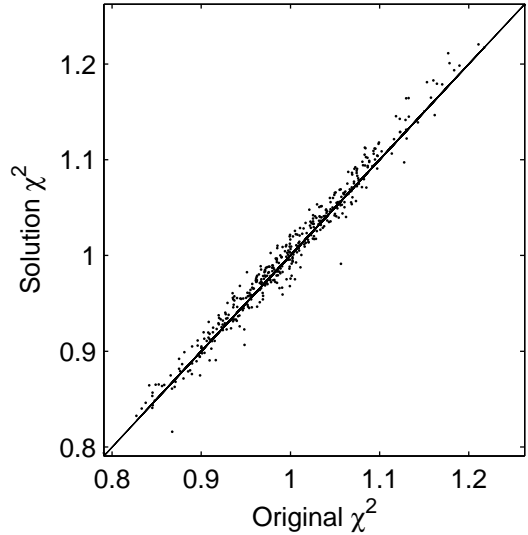}
\caption{Simulation results: the derived vs. original {\it normalized}
$\chi^2$.}
\label{fig:simulation2_rms}
\end{figure*}

To assess the goodness-of-fit of the solutions we calculated for each
system the normalized $\chi^2$ of the EBAS solution, which is the sum
of squares of the residuals, scaled by the uncertainty of each point,
divided by the number of degrees of freedom of each solution. We
compared this value with the ``original $\chi^2$'' of each lightcurve,
which is the average of the sum of squares of the inserted errors
around the original calculated lightcurve, again scaled by the
uncertainty of each point.  Fig~\ref{fig:simulation2_rms} shows the
$\chi^2$ of the solution versus the original one.  The continuous line
is the locus of points for which the two $\chi^2$s are equal. The
figure shows that most points lie next to the line, which means that
for each lightcurve the algorithm found a set of parameters that fit
the data with residuals which are close, on the average, to the
original scatter. While one can not be sure that global minima were
found for all lightcurves, the fact that none of the solutions showed
substantially large normalized $\chi^2$ is reassuring.

\begin{figure*}
\includegraphics{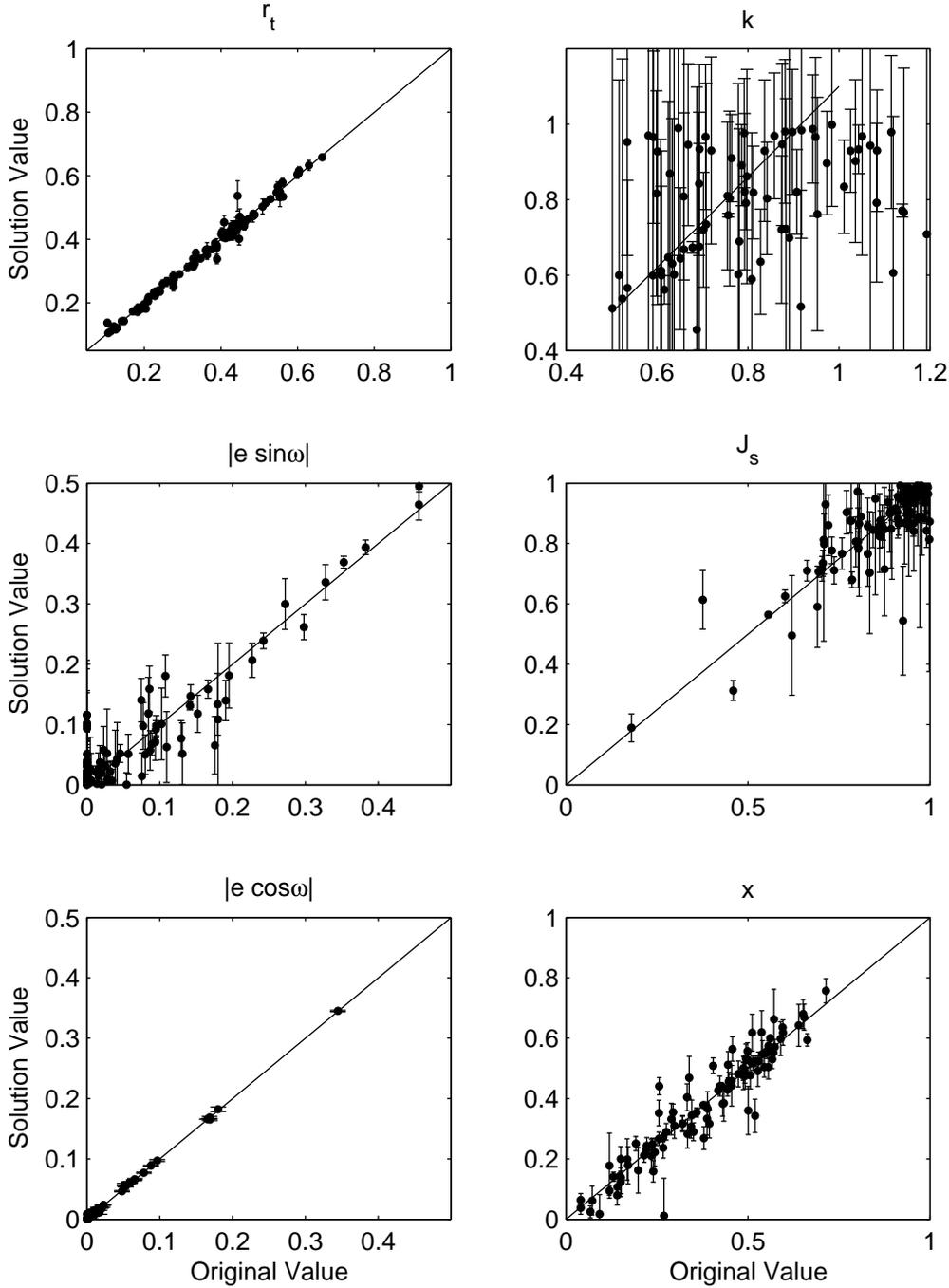}
\caption{Simulation results: the derived vs. ``true'' elements.}
\label{fig:simulation2}
\end{figure*}

Fig~\ref{fig:simulation2} shows the values of six of the derived
elements of the simulated sample as a function of the original
values. In order not to turn the plot too dense, we randomly choose only
100 systems for the display.  The figure shows that the sum of radii,
$r_t$, is reproduced quite well by the code, and so is $e\cos\omega$. For
$x$,  $e\sin\omega$  and $J_s$, EBAS produced slightly less accurate,
but still quite good results. The parameter $k$ seems more difficult
to determine, and its derived values coincide with the original ones
only for lightcurves of high $S/N$ ratio. Still, the correlation
between the original and derived $k$ values for the whole sample is
0.6, and we feel that allowing $k$ to vary is meaningful, except
perhaps for very noisy lightcurves.

\begin{figure*}
\includegraphics{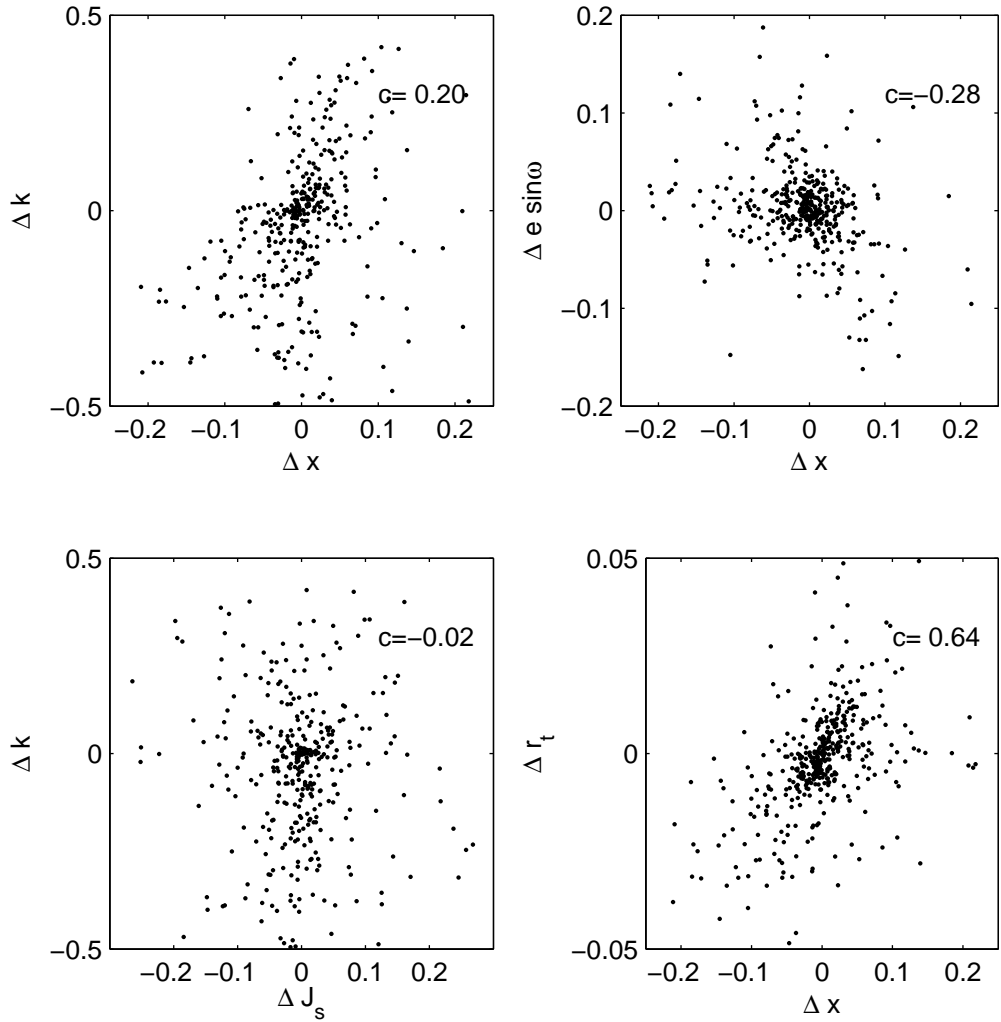}
\caption{Correlation between the deviations from true values for four
pairs of elements. The value of the corresponding correlation appears
in each panel.}
\label{fig:sim_corr}
\end{figure*}

It is well known that lightcurves of only one colour include
degeneracy between few parameters. The values of those parameters
deviate together from their true values, yielding almost as good
solutions as the ones with the true values. To estimate the magnitude
of this effect, we consider the deviations of the derived elements
from their true values in our simulations and estimate the
correlations between those deiviations.  Fig~\ref{fig:sim_corr} shows
the correlation between the deviation of the $x$ parameter --- $\Delta
x$, and three other parameter deviations.  The figure shows a small
but somewhat significant correlation with $\Delta e\sin\omega$ and
$\Delta k$, and high correlation with $\Delta r_t$. However, as the
deviations of most of the $r_t$ values are quite small, we still
suggest that the derived values of $r_t$ are valid. The figure also
shows that there is no correlation between $\Delta J_s$ and $\Delta
k$.

\begin{figure*}
\includegraphics{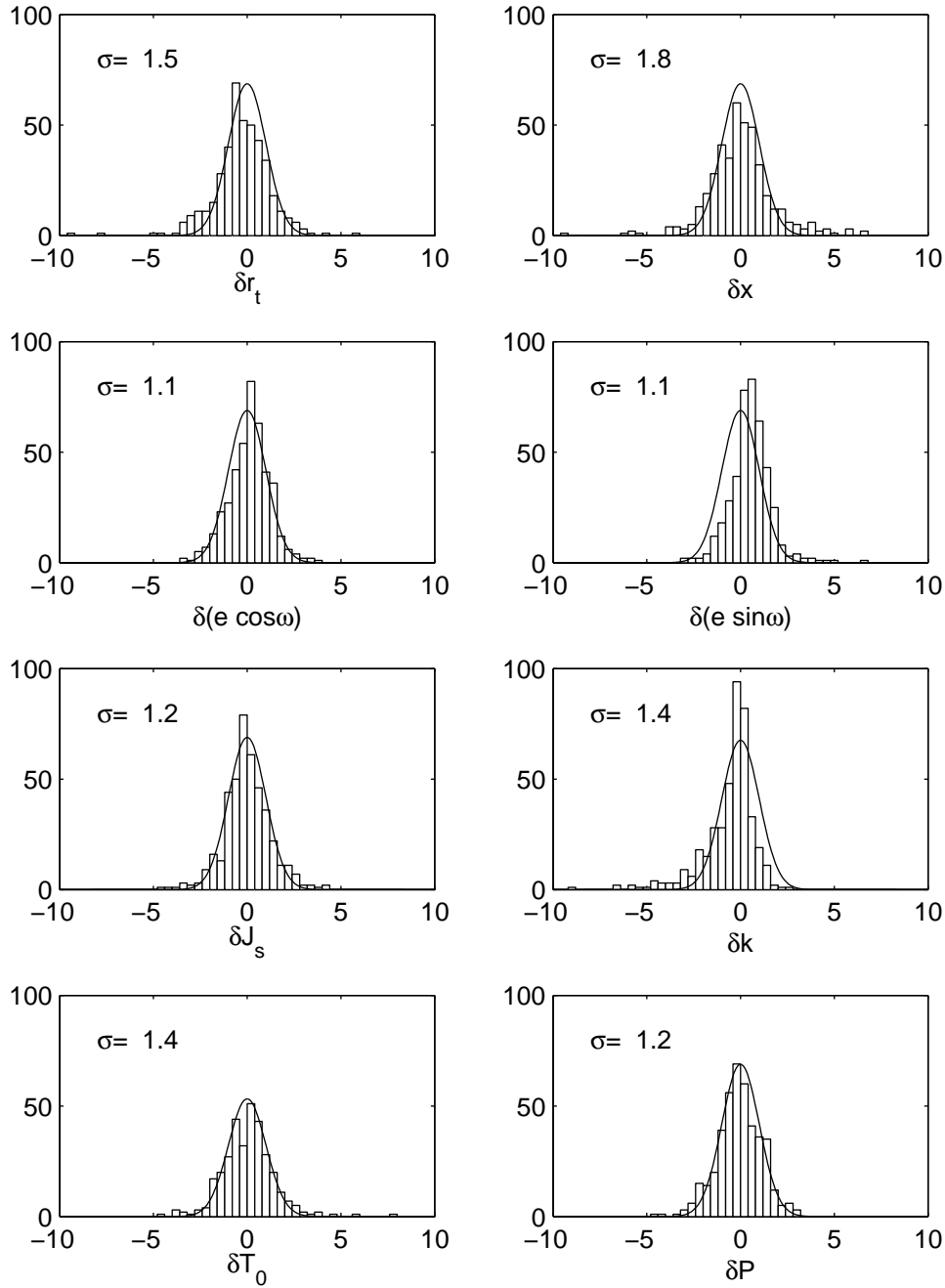}
\caption{Simulation results: the distribution of the scaled errors
for eight EBAS parameters.  For
comparison, Gaussian distributions with unity variance are
plotted. The rms of each distribution is given in each panel.}
\label{fig:simulation_err}
\end{figure*}

To explore the reliability of our uncertainty estimate we consider for
each parameter $p$ the scaled error $\delta p=
(p_{derived}-p_{original})/{\sigma_p}$, which measures the actual
error, i.e. the difference between the derived and original values of
$p$, divided by the uncertainty, $\sigma_p$, as estimated by EBAS. We
plotted in Fig~\ref{fig:simulation_err} histograms of scaled errors
for eight parameters. With correct uncertainties, the distributions of
the scaled errors should all have Gaussian shape and variance of
unity. Wide distribution indicates that our estimate for the error
might be too small. We can see that all distributions --- except that
of $\delta k$, the most problematic parameter --- are close to have a
Gaussian shape and width of unity, even though asymmetry and outliers
increase the rms value by up to a factor of two.

\subsection{The sensitivity of the elements to two simplifying
assumptions}

\begin{figure*}
\includegraphics{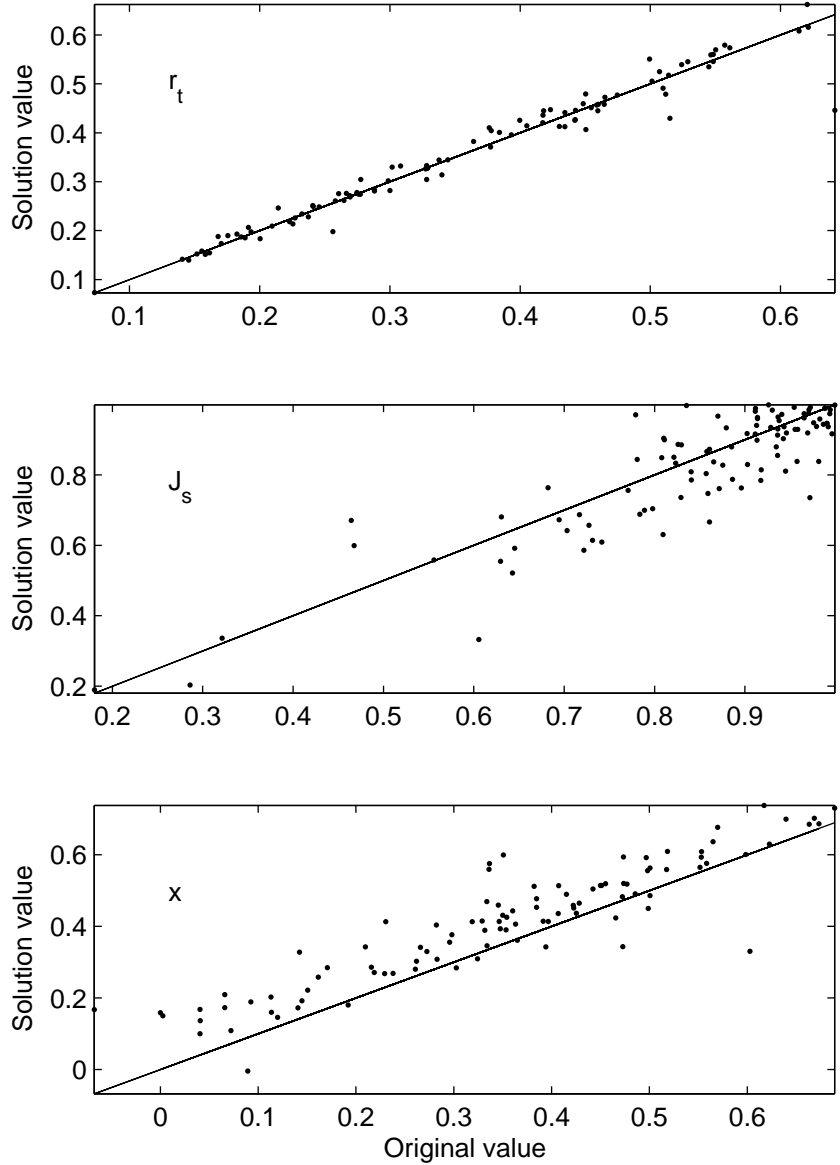}
\caption{Simulation results: the effect of the assumption $L_3=0$.  The
derived values of the parameters, assuming $L_3=0$ vs. the true values
for systems with $L_3=0.1$.}
 \label{fig:L3_0.1}
\end{figure*}

The present embodiment of EBAS assumes that $L_3$ is zero and the mass
ratio is unity. The former assumption implies that all the light of
the system is coming from the two components of the binary. This is
not necessarily the case, as a third star, either a background star or
a distant companion of the system, could also contribute to the total
light of the system. Failure to realize the contribution of a third
star could result in underestimation of the depth of the eclipses,
which induces further systematic errors in the derivation of the
binary elements. To estimate the error induced by the value assigned
to the light of a third star, we generated lightcurves identical to the
ones of the previous simulation, except that we set $L_3=0.1$ for all
of them. We then solved them using EBAS as before, assuming
$L_3=0$. The comparison between the solutions and the original values
for three elements --- the sum of radii, the impact parameter and the
ratio of radii, is plotted in Fig~\ref{fig:L3_0.1}.

The sum of radii, which is mainly sensitive to the eclipse {\it
shape}, is almost not affected by the different value of $L_3$. On the
other hand, the values of the surface brightness ratio show a
relatively large spread relative to the ``true'' values. However, this
spread is not larger than the corresponding one in
Fig~\ref{fig:simulation2}. This means that the assumption $L_3=0$
did not increase substantially the error of the derived values of the
surface brightness ratios. The impact parameter values show the
clearest effect. The derived values are systematically larger than the
true values, in order to account for the shallower eclipses
interpreted by EBAS, because of the $L_3=0$ assumption.

\begin{figure*}
\includegraphics{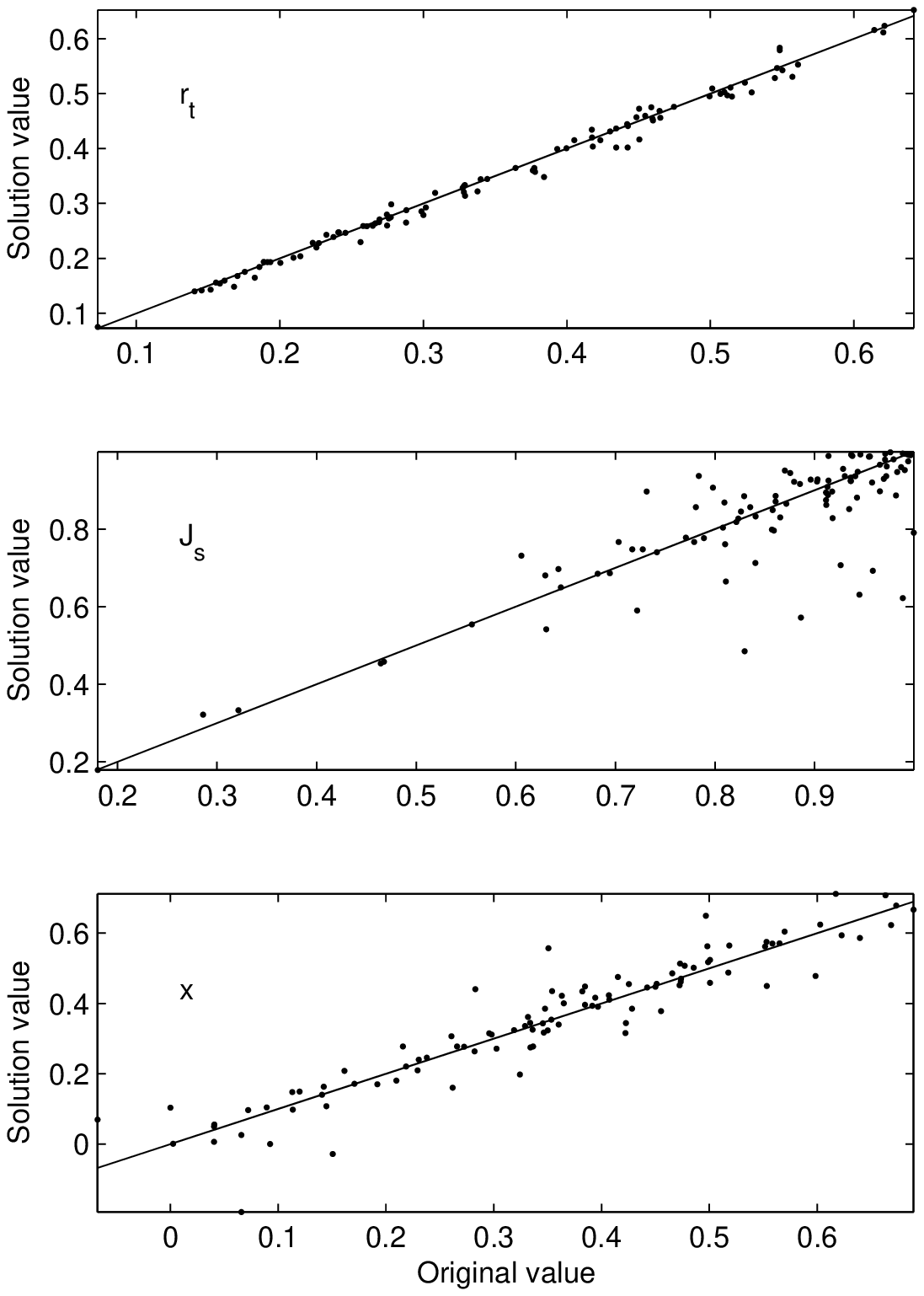}
 \caption{Simulation results: the effect of the assumption that the
     mass ratio is unity. The derived values of the parameters,
     assuming $q=1.0$ vs. the true values for systems with $q=0.8$.}
 \label{fig:Q_0.8}
\end{figure*}

We performed similar simulation to estimate the effect of the
assumption that $q=1$. The results are plotted in Fig~\ref{fig:Q_0.8}.
The simulations show that the assumption of $q=1$ does not affect
substantially the derived values of the sums of radii, the surface
brightness ratios and the impact parameters.

\subsection{The real OGLE LMC lightcurves --- comparison with 
manual solutions}

As another test of EBAS, we applied our algorithm to the OGLE LMC
lightcurves solved by NZ04 using manual iterations with EBOP.  Note
that we compare here the ``manual'' fits by NZ04 with those of EBAS
for the real systems, while Fig~\ref{fig:simulation2_rms} compares the
original scatter of {\it simulated} lightcurves with that
resulting from the fit.

\begin{figure*}
 \includegraphics{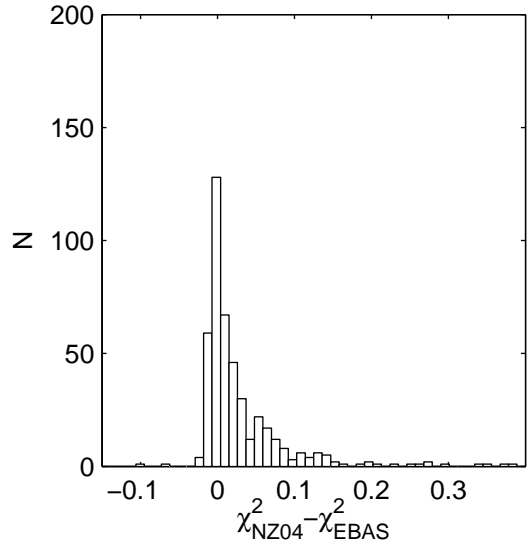}
 \caption{Histogram of difference between NZ04 solutions $\chi^2$ and EBAS solutions $\chi^2$.}
 \label{fig:NZ_rms}
\end{figure*}

After discarding 58 solutions with high alarm or high $\chi^2$, we
were left with 451 binaries. To compare EBAS solutions with those of
NZ04, we derive for each EBAS solution a normalized
$\chi^2$, which is equal to the unnormalized one, given by
Eqn.~\ref{chi2}, divided by the number of observed points minus the
number of fitted parameters.  Fig~\ref{fig:NZ_rms} plots a histogram of
the NZ04 normalized $\chi^2$'s minus those of EBAS.

The comparison shows that the two sets of solutions are comparable. In
fact, NZ04 achieved better solutions for 156 systems, out of which
only 2 systems, which can be seen in the figure, had smaller
normalized $\chi^2$ by more than 3\%. On the other hand, EBAS solved
295 systems with lower $\chi^2$, out of which 156 solutions had
smaller normalized $\chi^2$ by more than 3\%.  We therefore suggest
that EBAS found slightly better solutions for most of the binaries
analysed by NZ04.

\subsection{Four eclipsing binaries analysed by GOMoM05 ---
comparison with the WD solutions}

\begin{figure*}
\includegraphics{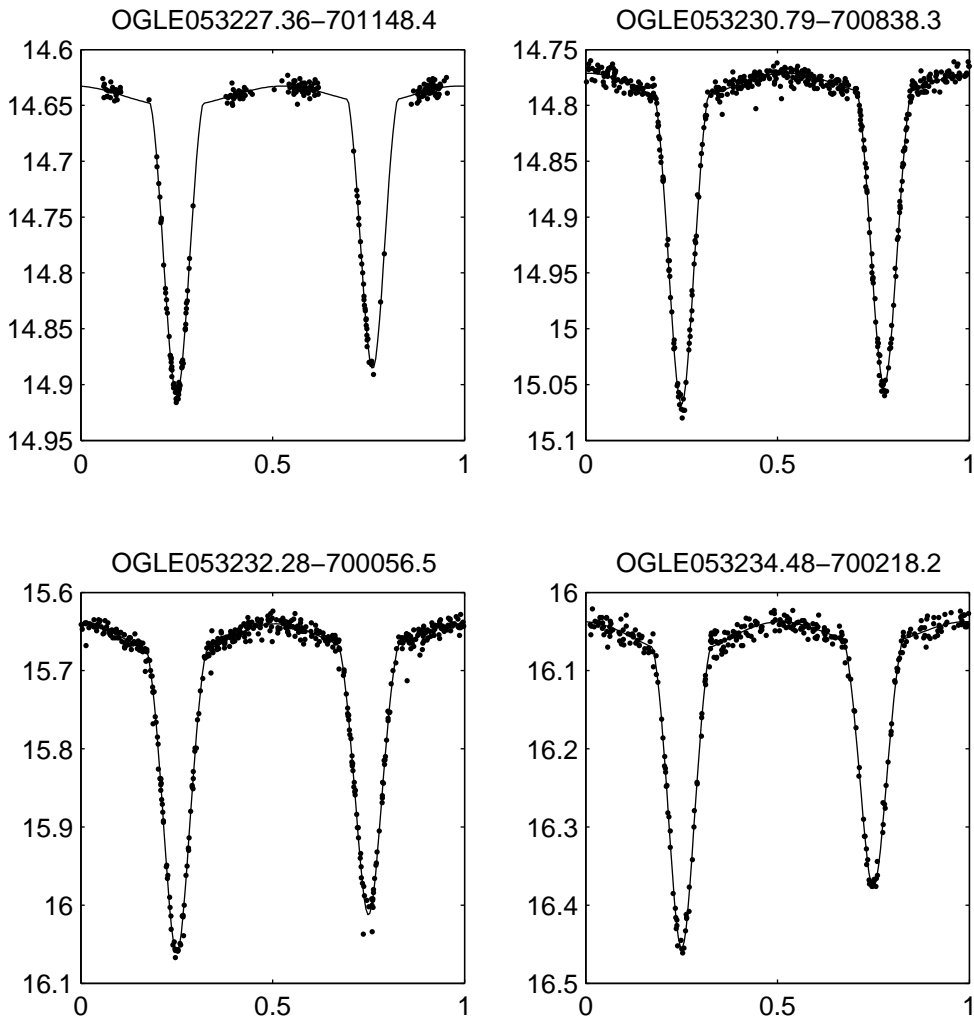}
 \caption{The EBAS solutions for the OGLE lightcurves of the four
 binaries analysed by GOMoM05.}
 \label{fig:four_binaries}
\end{figure*}

Very recently, GOMoM05 derived absolute parameters for eight eclipsing
binaries in the LMC, using photometric data from MACHO
\citep{alcocketal97} together with a few radial-velocity
measurements. OGLE data is available for four of these systems, and
GOMoM05 used these data as well. To compare the values of GOMom05 with
EBAS, we solved for these four systems and plotted their solutions in
Fig~\ref{fig:four_binaries}.

Before comparing the results of the two solutions, a word of caution
is needed. GOMoM05 used the WD code, derived the temperature ratio
from the spectroscopic data, and used lightcurves of three different
colours for each of the four systems. Our solution is based on the
OGLE $I$-band data only.  We therefore choose to compare only the
geometric parameters of the systems, namely the sum of radii, the
inclination, the ratio of radii and the eccentricity.
Table~\ref{tab:four_binaries_comparison} brings the detailed
comparison. For each of the four systems, the first line in the table
gives the GOMoM05 elements, while the second line gives EBAS's. It is
reassuring that despite all the differences in the derivation of the
two sets of elements, all values of all geometric elements agree
within $1$--$2$ $\sigma$ of each other. Indeed, the large differences
in the values of the ratio of radii of OGLE051804.81-694818.9 and
OGLE052235.46-693143.4 are only caused by a switch between the primary
and the secondary in the GOMoM05 solution. The reciprocal GOMoM05's
values are within 1$\sigma$ of the EBAS results.

\begin{table*}
\label{tab:four_binaries_comparison}
\centering
\begin{minipage}{140mm}
\caption{Elements of four binaries: comparison between GOMoM05 (first row) and EBAS (second row) solutions. }
\begin{tabular}{@{}lcccc@{}}
System & $r_t$ & $i$ & $k$ & $e$\\
\hline
OGLE052232.68-701437.1 & $   0.426 \pm    0.018$ & $    78.0 \pm      0.8$ & $    0.84 \pm     0.10$ & $   0.025 \pm    0.006$ \\
& $   0.458 \pm    0.007$ & $    77.0 \pm      0.4$ & $    0.99 \pm     0.07$ & $   0.044 \pm    0.014$\\
OGLE050828.13-684825.1 & $   0.458 \pm    0.009$ & $    77.9 \pm      0.8$ & $    0.80 \pm     0.20$ & $   0.043 \pm    0.006$ \\
& $   0.475 \pm    0.003$ & $    77.3 \pm      0.2$ & $    0.84 \pm     0.12$ & $   0.043 \pm    0.001$\\
OGLE051804.81-694818.9 & $   0.498 \pm    0.010$ & $    81.3 \pm      0.8$ & $    0.75 \pm     0.08$ & 0\\
& $   0.496 \pm    0.004$ & $    81.0 \pm      0.3$ & $    1.49 \pm     0.04$ & $   0.003 \pm    0.004$\\
OGLE052235.46-693143.4 & $   0.485 \pm    0.017$ & $    80.1 \pm      1.4$ & $    0.73 \pm     0.09$ & 0\\
& $   0.490 \pm    0.007$ & $    80.4 \pm      0.8$ & $    1.49 \pm     0.25$ & $   0.015 \pm    0.010$\\
\hline
\end{tabular}
\end{minipage}
\end{table*}

\section{Discussion}            %
\label{sec:discussion}

We have shown that it is possible to solve lightcurves of eclipsing
binaries with a fully automated algorithm which is based on the EBOP
code. Our simulations have shown that the results of EBAS are close to
the ``real'' ones and that EBAS results for most cases have a quality
which is better than is achieved with human interaction.

Although the EBOP code does not include the sophistications offered
by, e.g., the widely used Wilson-Devinney programme, it has the
advantage of being simple and of producing parameters closely related
to the real information content of the lightcurve (e.g., surface
brightness instead of effective temperature). In addition, it does
take into account not only reflection effects, but also tidal
deformation of components (even though in a primitive way), so that it
remains useful for systems with moderate proximity effects. Comparison
with the recent work of GOMoM05 who used the WD code to analyse three
colour photometry, radial-velocity and spectroscopic data of four
systems shows that the present version of EBAS recovers quite well the
sum of radii, the inclination and the ratio of radii.

It is interesting to compare the speed of the present version of EBAS
with the very fast automated DEBiL algorithm \citep{devor2005}, which
was used to derive the elements of almost 10,000 eclipsing binaries in
the Galactic bulge.  On the average, it took EBAS 50 seconds CPU time to solve
one orbit on an AMD Opteron, 250 2.46GHz, 64-bit machine, while error
estimation took another 100 seconds. This is about 3 times longer than
it took Devor to solve an orbit with his DEBiL with a SUN UltraSPARC5
333MHz. Applying EBAS to 10,000 systems is therefore feasible.

EBAS uses two redundant techniques to ensure the finding of the global
minimum --- a search for the minimum with simulated annealing and
consultation with the new alarm $\mathcal A$. The simulated annealing
technique causes heavy computation load on EBAS, and if the number of
lightcurves is too large, the demanding parameters of the annealing
can be slightly relaxed, since we can rely on the alarm to warn us if
the global minimum is not reached.

In the next papers we plan to apply EBAS to the sample of the LMC
(Mazeh, Tamuz \& North 2005, Paper II) and SMC OGLE data. Obviously,
when EBAS is applied to real data, one should carefully examine the
implication of the specific choices done for the nonvariable
parameters, the values of the mass ratio, the fractional light of a
possible third star and the values for the limb and gravity
darkening. However, the goal of applying EBAS to such large datasets
is not to derive the exact parameters of a particular system. Instead,
the aim is to study statistical features of the short-period binaries,
like their frequency and period distribution. In that sense, the set
of data points we use includes the photometry obtained for all the
eclipsing binaries found in the sample. For the OGLE LMC data, this is
about 300 points for more than 2000 systems, which adds up to about
0.6 millions points, admittedly with low $S/N$ ratio. Such a huge
dataset should allow us to study some statistical features of the
short-period binaries.

An obvious extension of EBAS would be to allow for automated
derivation of the mass ratio, the light of a third star, and even the
limb and gravity darkening. This can not be done with OGLE data of the
LMC, but would be possible for systems with better data and more than
one colour photometry. For close binaries with strong proximity
effects we plan to allow EBAS to use the WD code. In principle, the
approach should be the same, with the same procedure to find the best
initial guess, the same simulated annealing search and the same error
estimation. The development of these capacities of EBAS are
deferred to a later paper.

Finally, we plan to construct an automated algorithm to derive the
masses of the two stars in each eclipsing binary in the LMC, in a
similar approach to the one presented by J. Devor in the ``Close
Binaries in the 21st Century: New Opportunities and Challenges''
meeting. Our approach relies on the fact that we know the distance to
all the binaries in our neighbouring galaxy, up to a few percent, and
therefore know the absolute magnitude of the LMC OGLE systems. OGLE
data includes some measurements in the $V$ band for each star in the
LMC, and therefore the available absolute magnitude information
includes two colours. Furthermore, MACHO data \citep{alcocketal97} is
also available for most of these systems. This should suffice to
derive a crude estimate of the masses and ages of all the eclipsing
binaries in the OGLE LMC dataset.

\section*{Acknowledgments}
We are grateful to the OGLE team, and to L. Wyrzykowski in particular,
for the excellent photometric dataset and the eclipsing binary
analysis that was made available to us. We thank J. Devor, G. Torres
and I. Ribas for very useful comments. The remarks and suggestions of
the referee, T. Zwitter, helped us to substantially improve the
algorithm and the paper. This work was supported by the Israeli
Science Foundation through grant no. 03/233.

\bibliographystyle{mn2e}
\bibliography{ref}
\end{document}